\documentclass[twocolumn,superscriptaddress,amsmath, amssymb, amsfonts,preprintnumbers,aps,prd,longbibliography,nofootinbib
]{revtex4-2}

\usepackage{graphicx}
\usepackage{dcolumn}
\usepackage{bm}
\usepackage{tikz,tikz-feynman}
\usepackage{cleveref}
\crefname{section}{Section}{Sections}
\crefname{appendix}{Appendix}{Appendices}
 
\newcommand{\pb}[1]{\left\{#1\right\}\textsubscript{PB}}
\newcommand{\cQ}{\mathcal{Q}}
\newcommand{\cH}{\mathcal{H}}

\newcommand{\cS}{\mathcal{S}}
\begin{document}

\preprint{MS-TP-24-35, TUM-HEP-1564/25}

\title{Higher-spin effects in black hole and neutron star binary dynamics: \\worldline supersymmetry beyond minimal coupling
}

\author{Domenico Bonocore}
 \email{domenico.bonocore@tum.de}
\affiliation{
TUM School of Natural Sciences,
Technische 
	Universit\"at M\"unchen, 
	Physik Department,
	James-Franck-Straße 1, D-85748, 
	Garching, Germany
}

\author{Anna Kulesza}
 \email{anna.kulesza@uni-muenster.de}
\affiliation{ Institut f\"{u}r Theoretische Physik, Universit\"{a}t M\"{u}nster, Wilhelm-Klemm-Stra\ss e 9,
	D-48149 M\"{u}nster, Germany}

\author{Johannes Pirsch}
\email{johannes.pirsch@uni-muenster.de}
 \affiliation{ Institut f\"{u}r Theoretische Physik, Universit\"{a}t M\"{u}nster, Wilhelm-Klemm-Stra\ss e 9,
	D-48149 M\"{u}nster, Germany}

\date{\today}

\begin{abstract}
The inclusion of spin effects in the binary dynamics for black hole and neutron stars is crucial 
for the computation of gravitational wave observables.
Worldline supersymmetric models have shown to be particularly efficient at this task up to quadratic order in spin,
but progress at higher orders has been hampered by no-go-theorems. 
In this work we propose a novel approach to overcome this problem by extending the supersymmetry beyond minimal coupling. We demonstrate the potential of this approach by computing an all-order in spin and linear in curvature, manifestly supersymmetric Hamiltonian, as well as a cubic order in spin Hamiltonian in arbitrary spacetime dimensions. In doing so, we identify a criterion that uniquely determines the Kerr geometry 
in terms of worldline supersymmetry. 
Equipped with these Hamiltonians, we demonstrate the exponentiation of three-point and Compton amplitudes 
using the recently proposed Generalized Wilson line approach.
\end{abstract}

\maketitle


\section{Introduction}
A fascinating aspect of future gravitational wave astronomy lies in the ability to infer fundamental properties of black holes and neutron stars from the detected waveforms. 
The upcoming third generation of gravitational wave detectors will enable reaching an unprecedented precision for observable measurements \cite{Punturo:2010zz, Abac:2025saz, LISA:2017pwj, LISA:2022yao, LIGOScientific:2016wof, Chamberlain:2017fjl, Ballmer:2022uxx}, among them spin and tidal deformability. Such progress requires accurate waveform templates that in particular capture spin effects in the binary dynamics. 

A wealth of analytical results have been derived in the Post-Minkowskian (PM) expansion, i.e.\ the expansion around weak gravitational fields in Newton's constant $G$. The driving force behind this breakthrough is largely due to a modern formulation of the two-body problem in general relativity (GR) with quantum field theory (QFT) techniques \cite{Damour:2017zjx, 2019binaryblackholes,Mogull:2020sak,Bern:2020buy,Neill:2013wsa,Luna:2017dtq,Kosower:2018adc,Cristofoli:2021vyo,Bjerrum-Bohr:2013bxa,Bjerrum-Bohr:2018xdl,Bern:2019nnu,Bern:2019crd,Bjerrum-Bohr:2021wwt,Cheung:2020gyp,Bjerrum-Bohr:2021din,DiVecchia:2020ymx,DiVecchia:2021bdo,DiVecchia:2021ndb,DiVecchia:2022piu,Heissenberg:2022tsn,Jakobsen:2022psy,Jakobsen:2022fcj,Bellazzini:2022wzv,Bastianelli:2021nbs,Damour:2020tta,Herrmann:2021tct,Damgaard:2019lfh,Damgaard:2021ipf,Damgaard:2023vnx,Aoude:2020onz,AccettulliHuber:2020dal,Brandhuber:2021eyq,Bern:2021dqo,Bern:2021yeh,Bern:2022kto,Bern:2023ity,Damgaard:2023ttc,Brandhuber:2023hhy,Brandhuber:2023hhl,DeAngelis:2023lvf,Herderschee:2023fxh,Caron-Huot:2023vxl,FebresCordero:2022jts,Bohnenblust:2023qmy,Bern:2024adl,Alaverdian:2024spu,Akpinar:2024meg,Bohnenblust:2024hkw,Jakobsen:2023ndj,Jakobsen:2023hig,Driesse:2024xad,Foffa:2019hrb,Foffa:2019rdf,Foffa:2019yfl,Adamo:2024oxy,Akpinar:2025bkt,Akpinar:2025huz} together with advances in integration methods for Feynman integrals \cite{Beneke:1997zp,Gehrmann:1999as,Laporta:2000dsw,Smirnov:2008iw,Henn:2013pwa,Klappert:2020nbg,Dlapa:2020cwj,Gardi:2024axt,delaCruz:2024xsm}. Importantly for this program, 
an effective two-body potential governs both the dynamics of the unbound scattering scenario and dynamics of an inspiraling binary,
a connection that takes a particularly elegant form in the so-called bound-to-boundary map
\cite{Kalin:2019rwq,Kalin:2019inp,Cho:2021arx}.

In this QFT-inspired approach, the rotation of black holes and neutron stars is represented as the spin $s$ of a point particle, while finite size effects are captured using effective operators modeling the internal structure of the objects \cite{Goldberger:2004jt,Levi:2015msa,Kalin:2020mvi,Bern:2020buy}. 
Correspondingly, apart from the expansion in $G$, one should also consider an expansion in the macroscopic spin variable $S$, defined as the conjugate momentum of the angular velocity. The current state of the art in the $G^nS^m$ double expansion is the combined $n+m$=5 accuracy \cite{Driesse:2024xad,Bern:2024adl,Jakobsen:2023ndj,Jakobsen:2023hig,Bern:2022kto,Bern:2021dqo,Jakobsen:2022fcj,FebresCordero:2022jts,Bohnenblust:2024hkw,Akpinar:2024meg}. Additionally, considerable progress has been made at higher orders in spin \cite{Chung:2020rrz,Bohnenblust:2024hkw,Aoude:2022thd,Alessio:2022kwv,Damgaard:2022jem,Chen:2021kxt,Bautista:2021wfy,Bautista:2022wjf,Guevara:2018wpp,Cheung:2023lnj,Ben-Shahar:2023djm,Haddad:2024ebn,Akpinar:2025bkt,Bjerrum-Bohr:2025lpw}, where for example results are available for the sectors 
$\mathcal{O}(GS^\infty)$, $\mathcal{O}(G^2S^\infty)$ and $\mathcal{O}(G^3S^4)$.
 In fact, GR provides an upper bound for the spin of black holes, via $S\leq Gm^2$, 
 while neutron stars seem to be restricted to slower rotations \cite{Cipolletta:2015nga}. 
Considering this scaling with $G$, higher orders in spin are elevated to the same level of importance as higher orders in the PM expansion, and are thus of primary relevance for improving the theoretical precision.

While the need to model spin effects is evident, a QFT description of these classical 
effects is not without challenges. Taking the
classical limit in the scattering amplitude approach is quite involved \cite{Kosower:2018adc,Maybee:2019jus,Capatti:2024bid}
and the identification of the particle spin $s$ with the macroscopic spin variable $S$ requires interacting higher-spin fields \cite{Bern:2020buy}. Worldline models improve on these aspects considerably by efficiently reorganizing classical parts of QFT amplitudes 
into terms that have a well-defined $\hbar\to0$ limit \cite{Mogull:2020sak,Jakobsen:2021zvh,Bonocore:2021qxh,Bonocore:2024uxk,Ajith:2024fna,Damgaard:2023vnx}.
Crucially, the worldline approach enables to represent
particles of spin $s=\frac{1}{2}$ or $s=1$ minimally coupled to gravity with a ${\cal N}=2s$ supersymmetric action \cite{Brink:1976uf,Brink:1976sz,Henneaux:1982ma}, capturing spin effects through $O(S^{2s})$, most prominently explored in the worldline QFT (WQFT) approach \cite{Jakobsen:2021zvh,Jakobsen:2022fcj,Jakobsen:2023ndj,Jakobsen:2023hig} and more recently using generalized Wilson lines (GWLs) \cite{Bonocore:2024uxk}. The identification of the classical spin variables is particularly clear in the GWL approach, thanks to the synergy of the supersymmetric formulation and the soft expansion \cite{Bonocore:2021qxh, Bonocore:2024uxk}.
Extending a worldline description beyond quadratic order in spin, however, has proven challenging although promising progress has recently been reported \cite{Vines:2016unv, Haddad:2024ebn, Ben-Shahar:2023djm}.
Supersymmetric formulations, in particular, are hindered by well-known no-go theorems \cite{Howe:1988ft,Bastianelli:2015tha,Bastianelli:2007pv,Bekaert:2010hw}. One possibility which has recently been explored in \cite{Haddad:2024ebn} is to abandon supersymmetry in favour of bosonic fields for the spin tensor. In this work we approach the problem from a different angle by exploring non-minimal coupling on the supersymmetric worldline.

Non-minimal coupling has a long and extensive history. It emerges
straightforwardly in the context of effective field theories (EFTs), where higher dimensional operators are added to the action to model physics below a certain energy scale, while the UV physics is absorbed into Wilson coefficients.
For classical gravity, in particular, worldlines EFTs are formulated in terms of operators containing arbitrary powers of the curvature tensor, which therefore couple the point particle to the metric non-minimally. As such, they represent deviation from geodesic motion, and are thus associated with the finite extent of the source \cite{Goldberger:2004jt,Levi:2015msa,Kalin:2020mvi,Porto:2016pyg,Liu:2021zxr,Goldberger:2022ebt}. The operators that contribute at lowest order in the PM expansion are self-induced tidal effects. They correspond to the multipole moments of the object's stress-energy tensor that couple the spin $S$ of the particle to derivatives of the curvature. For spinning black holes the numerical value of the corresponding Wilson coefficients is unity, while values different from one capture the complicated internal structure of neutron stars instead.

A fresh look at non-minimal coupling comes from recent developments in on-shell methods for scattering amplitudes.
Specifically, in \cite{Arkani-Hamed:2017jhn} on-shell amplitudes for massive particles of arbitrary spin were introduced. While massless amplitudes exhibit a unique structure that corresponds to the usual notion of minimal coupling in QFT, massive amplitudes admit several distinct tensor structures. This suggests that massive higher-spin interactions naturally lead to higher dimensional operators. One of these structures is special and matches the properties of the massless minimally coupled amplitude in the high energy limit. In fact, it has been shown in \cite{Guevara:2018wpp,Chung:2018kqs} that these minimally coupled higher-spin amplitudes reproduce precisely the multipole moments that uniquely define the Kerr black hole geometry \cite{Scheopner:2023rzp,Levi:2015msa}. 
These multipole moments have a representation on the worldline as corrections to the dynamical mass of the particle. We will show in the following how they naturally emerge from the supersymmetry algebra. 

In this letter we propose a non-minimally coupled supersymmetric worldline model which describes the gravitational interaction beyond quadratic order in spin. After introducing non-minimal coupling, in section \ref{sec:wlsusy} we enforce the supersymmetry algebra order by order in spin, 
leading to a formal expression for the higher-spin Hamiltonian. Following that, in section \ref{sec:hamiltonians} we identify the subclass of non-minimal couplings that define the Kerr geometry on the worldline, enabling to provide for the first time two higher-spin Hamiltonians: A) the all-order-in-spin and linear-in-curvature Hamiltonian describing the dynamics of Kerr black holes in four spacetime dimensions and B) the cubic-in-spin Hamiltonian for general extended bodies in generic spacetime dimensions $d$. The $d$-dimensional expressions are interesting from a theoretical perspective \cite{Gambino:2024uge,Cristofoli:2020uzm,Bianchi:2024shc,Bianchi:2025xol} and computationally useful for constructing amplitude integrands, making them free of Levi-Civita tensors and therefore unambiguous in dimensional regularization.
In section \ref{sec:scattering} we demonstrate, using the GWL approach, that these Hamiltonians lead to the exponentiation of the higher-spin three-point and Compton amplitudes, which are the essential building blocks for the computation of observables.

\section{Spin through worldline supersymmetry}\label{sec:wlsusy}
Free massless particles of spin $s$ are characterized by $2s+1$ conserved charges on phase space. These are the $2s$ Grassmann valued charges $\cQ_i$ for $i\in\{1,\dots,2s\}$ and the Hamiltonian $\cH$ related to the charges via a $\mathcal{N}=2s$-fold supersymmetry algebra under Poisson brackets (PB) \begin{align}
    \pb{\cQ_i,\cQ_j}=&-2i\delta_{ij}\cH\label{eq:susy1}~.
\end{align}
The supersymmetry algebra implements a worldline analogue of the $2s$-fold tensor product of the Clifford algebra written in terms of momenta $\{p_\mu\gamma^\mu,p_\nu\gamma^\nu\}=2p^2$.
Minimal coupling to gravity is achieved by using the charge 
\begin{equation}
    \cQ_i=i\psi_i^a e_a^\mu \pi_\mu\equiv i\psi_i^a e_a^\mu \left(p_\mu+\frac{i}{2}\omega_\mu^{bc}\psi_b^i\psi_c^i\right),
\end{equation}
where $p_\mu$ is the canonical momentum, $\omega_\mu^{ab}$ is the spin connection, $e_a^\mu$ is the tetrad, and the Grassmann multiplet $\psi_i^a$ is related to the GR spin tensor $S^{\mu\nu}$ through 
\begin{equation}\label{eq:spintensor}
     S^{\mu\nu}=ie^\mu_a e^\nu_b \psi^a_i\psi^b_i.
 \end{equation}
This charge fails to obey the supersymmetry algebra for $s>1$ \cite{Howe:1988ft}. In order to describe particles of higher-spin we consider the non-minimally coupled charge 
\begin{equation}\label{eq:defq}
	\cQ_i=i\psi_i^a e_a^\mu P_\mu\equiv i\psi_i^a e_a^\mu (\pi_\mu+q_\mu)~,
\end{equation}
with yet to be determined $q_\mu$.
Correspondingly, 
\begin{align}
	\pb{\cQ_i,\cQ_j}=i\delta_{ij}P^2-\psi_i^a\psi_j^b T_{ab}~,
\end{align}
where the $q_\mu$ dependence is included in the new term $T_{ab}$ and $P^2$. 
Clearly, for vanishing $T_{ab}$ the brackets are diagonal and thus $\cal H$ can be defined via  \cref{eq:susy1} i.e.\ $\cH=-\frac{1}{2}P^2$.
The inclusion of a mass for the particle is trivial and simply changes the Hamiltonian to 
\begin{equation}\label{eq:dynamicalmass}
    \cH=\frac{1}{2}(m^2-P^2)\equiv \frac{1}{2}(\mathcal{M}^2-\pi^2),
\end{equation}
where in the last step we introduced the dynamical mass $\mathcal{M}$, 
which collects
all dependence over $q_\mu$.
The strategy we follow is 
therefore to solve $T_{ab}=0$ for $q_\mu$ order by order in curvature and spin,
hence enforcing supersymmetry 
up to the corresponding order in the expansions.
To do so, we first expand $q_\mu$ into a power series in the spin tensor
\begin{equation}\label{eq:fmu}
	q_\mu=\sum_{n=0}^{\infty}q_{\mu,(n)}S^{(n)},
\end{equation}
where $(n)$ is short for the $2n$ indices $a_1b_1\dots a_nb_n$.
Quite generally, the coefficients $q_{\mu,(n)}$  depend on the momentum $\pi_\mu$ and a collection of spacetime functions $\phi_I=\{R_{\mu\nu\varrho\sigma},\nabla_\alpha R_{\mu\nu\varrho\sigma},\dots \}$, where the ellipses denote higher orders in the derivatives and curvature tensor.
As such, they encode the coupling between the spin and the background spacetime. 

Since the $\mathcal{N}=2s$ model encodes the spin dynamics only to order $S^{2s}$ 
in the spin tensor, we can truncate the series in \cref{eq:fmu} at that order by requiring that the $q_{\mu,(n)}$ fulfil the Bianchi identity for any three indices $a_i,b_i,a_j$. This leads to the explicit constraint
\begin{align}\label{eq:constraintT}
	&T_{ab}=\frac{1}{2}R_{abcd}S^{cd}-\pb{q_a,q_b}\nonumber+R_{[a|cef}S^{ef}\frac{\delta q_{|b]}}{\delta \pi_c}\\&\nonumber-2\sum_{n=0}^{2s}S^{(n)}\Big(2(n+1)\pi^r q_{[a|,r|b](n)}-\nabla_{[a|}\phi_I \frac{\delta}{\delta \phi_I}q_{|b],(n)}\Big)\\
	&+2q_c e^\mu_f \omega_\mu^{dc}\frac{\delta}{\delta \pi_f}\eta_{d[b}q_{a]}+2 q_c \frac{\delta q_b}{\delta S_c^{~a}}-2 q_c \frac{\delta q_a}{\delta S_c^{~b}}=0.
\end{align}
In the following section we present two solutions to 
\cref{eq:constraintT}, which yield the corresponding higher-spin Hamiltonians of the supersymmetric model.

\section{Higher-spin Hamiltonians}\label{sec:hamiltonians}
\subsection{Linear curvature all order in spin in $d=4$}
Restricting our attention to the first order in curvature and $d=4$, we first define the spin vector $s_\mu$ and mass dipole $D_\mu$ in terms of the spin tensor of \cref{eq:spintensor} according to
\begin{equation}
    D_\beta+is_\beta\equiv r^\alpha S_{\alpha\beta}+i\frac{1}{2}r^\alpha\epsilon_{\alpha\beta\gamma\delta} S^{\gamma\delta},
\end{equation}
where the unit vector $r^\mu=\frac{\pi^\mu}{\sqrt{\pi^2}}$. Both $D_\mu$ and $s_\mu$ carry three degrees of freedom, due to orthogonality to $r_\mu$. Since the spin tensor captures spatial rotations, only three of the six degrees of freedom among $s_\mu$ and $D_\mu$ are physical. 
The unphysical ones are removed with 
an additional constraint called spin supplementary condition (SSC), which in 
the supersymmetric model is enforced as $\cQ_i=0$.
However, in order to maintain the supersymmetry of the action, all components of $s_\mu$ and $D_\mu$ are necessary at this stage. 

Demanding that $q_{\mu,(n)}$ is traceless, i.e.\ 
\begin{equation}\label{eq:Kerrcondition}
    q_{\mu,(n)}\eta^{a_ia_j}=0,
\end{equation}
the equation $T_{ab}=0$ dramatically simplifies and a solution for $q_\mu$ can be written down to all 
orders in $s_\mu$ and $D_\mu$. The resulting Hamiltonian then reads
\allowdisplaybreaks{
\begin{align}\label{eq:dynamicalmass_allorders}
	&\mathcal{H}=\nonumber \frac{m^2-\pi^2}{2}-\sum_{n=2}^{\infty}\sum_{k=0}^{n}\binom{n}{k}\frac{1}{n!|\pi|^{n-2}}(D^{b_i})^k(s^{b_j})^{n-k}\\
	&\times \left\{\begin{array}{ll}
		(-1)^{\frac{3n-k}{2}}E_{(b_1\dots b_{n-2},b_{n-1}b_n)}&\text{$n-k$ even}\\
		(-1)^{\frac{3n-k+1}{2}}B_{(b_1\dots b_{n-2},b_{n-1}b_n)}&\text{$n-k$ odd}\end{array}\right.
\end{align}}
where 
\begin{subequations}
\begin{align}
	E_{a_1\dots a_n,bc}=&\hat{\pi}^x\hat{\pi}^y\nabla_{a_1}\dots\nabla_{a_n}R_{xbyc},\\
	B_{a_1\dots a_n,bc}=&\frac{1}{2}\epsilon_{yczd}\hat{\pi}^x\hat{\pi}^y\nabla_{a_1}\dots\nabla_{a_n}R_{xbzd}.
\end{align}
\end{subequations}
Eq.~(\ref{eq:dynamicalmass_allorders}) is a new result, as it captures the all-order dependence on the
mass dipole, i.e.\ it does not rely on a specific choice of the SSC. 
Truncating \cref{eq:dynamicalmass_allorders} at order ${\cal O}(D)$ and making use of the differential Bianchi identity, 
we recover the result derived 
in \cite{Haddad:2024ebn}
 (see eq. (3.20) there, where the mass dipole is referred to as $Z_\mu$).

Importantly, this expression captures the multipole moments of Kerr black holes. In order to show this, we isolate the  $\mathcal{O}(D^0)$ term
\begin{align}\label{eq:fullconstraint_D0}
	\cH=\nonumber&\frac{m^2-\pi^2}{2}-\sum_{n=1}^{\infty}\frac{(-1)^n}{(2n+1)!} \left(\frac{s^a\nabla_a}{|\pi|}\right)^{2n-1} B_{ab}s^a s^b\\&-2\sum_{n=1}^{\infty} \frac{(-1)^n}{(2n)!} \left(\frac{s^a\nabla_a}{|\pi|}\right)^{2n-2}E_{cd}s^cs^d,
\end{align}
which agrees with the literature \cite{Levi:2015msa,Vines:2016unv,Scheopner:2023rzp}. 
We stress that the the Kerr geometry emerges from the conditions that
$q_{\mu,(n)}$ is traceless and fulfils the Bianchi identity. 

\subsection{Cubic in spin for generic $d$}
For generic spacetime dimensions $d$ and aiming at  non-Kerr solutions, we relax the traceless condition and make an ansatz for $q_{\mu,(n)}$ up to $n=3$ containing the curvature and its derivative. The resulting dynamical mass is conveniently expressed in terms of the mass dipole $D_\mu$ and the physical spin tensor $\tilde S_{\mu\nu}=S_{\mu\nu}-2r_{[\mu}D_{\nu]}$. For the $D_\mu=0$ SSC, terms that are quadratic in the mass dipole are irrelevant and we consider the linearized Hamiltonian
\allowdisplaybreaks{
\begin{align}\label{eq:Hamiltonian_d_wilsoncoefficients}
	&\cH=\nonumber\frac{m^2-\pi^2}{2}+\pi^2C_1 \tilde S_{ab}\tilde S^{ab}+\frac{1}{2}C_{D\hat{\pi}ab}\tilde S^{ab}\\\nonumber&-\frac{C_{ES^2}+\tilde C_{ES^2}}{8}\tilde S^{ab}\tilde S^{cd}\left(1-|\pi|^{-1}\nabla_D\right)C_{abcd}\\
	&\nonumber -\frac{\tilde C_{ES^2}}{2}\tilde S_{c}{}^{a} \tilde S^{cb}\left(1-|\pi|^{-1}\nabla_D\right)E_{ab}\\
	&\nonumber-\frac{C_{BS^3}}{6|\pi|}\nabla_a C_{\hat{\pi}bcd} \tilde S^{ad} \tilde S^{be}\tilde S^{c}{}_{e}-\frac{\tilde C_{BS^3}}{6|\pi|}\nabla_{\hat{\pi}}C_{aD\hat{\pi}b}\tilde S^{ac}\tilde S^{b}{}_{c}\\
	&+\frac{1}{|\pi|}\left(\tilde C_{ES^2}+\tilde C_{BS^3}\right)E_{a,bc}\tilde S^{ac} (D\cdot \tilde S)^b+\mathcal{O}(D^2).
\end{align}}
The coefficients $C_{ES^2}, \tilde C_{ES^2},C_{BS^3}$ and $\tilde C_{BS^3}$ are introduced as shorthands for combinations of coefficients coming from the ansatz and are identified such that in $d=4$ we recover the usual couplings in terms of $C_{ES^2}, C_{BS^3}$. This Hamiltonian is a new results and extends effective higher-spin Hamiltonians in the literature to generic spacetime dimensions. In particular, the Wilson coefficients $\tilde C_{ES^2}$ and $\tilde C_{BS^3}$ are new for generic $d$. They cancel or vanish in the special case $d=4$ 
\begin{align}
    	&\cH\overset{d=4}{=}\frac{m^2-\pi^2}{2}-2\pi^2 C_1 s^as_a +B_{Ds} \\&\nonumber+\frac{C_{ES^2}}{2}\left(1-|\pi|^{-1}\nabla_D\right)E_{ss}+\frac{C_{BS^3}}{6|\pi|}B_{s,ss}+\mathcal{O}(D^2),
\end{align}
which is in agreement with \cite{Marsat:2014xea,Haddad:2024ebn}.
Interestingly, the supersymmetric model also includes mass corrections \cite{Steinhoff:2014kwa} parameterized by the $C_1$ coefficient multiplying the constant $s^2$. In generic $d$ the Wilson coefficients are not fixed completely in the Kerr limit, but rather depend on a single unconstrained parameter $D$
\begin{subequations}
	\begin{align}
		C_{ES^2}=&\frac{d^2-4}{12}+Dd(d-4)(d+1),\\
		\tilde C_{ES^2}=&\frac{d+4}{12}+d(d+1)D,\\
		C_{BS^3}=&\frac{6}{2(d-1)},\\
		\tilde C_{BS^3}=&-\frac{1}{4}-3(d+1)D.
	\end{align}
\end{subequations}
We interpret this free parameter as an indication for a larger number of Kerr-like solution to Einstein's equation in $d>4$. As the uniqueness theorems do not apply there, we should not expect all multipole moments to be determined, which agrees with findings in \cite{Gambino:2024uge}. In $d=4$ however the Kerr limit fixes all coefficients to unity as expected for Kerr black holes.

\section{Wilson lines and exponentiation in gravitational scattering}\label{sec:scattering}

The classical scattering of compact astrophysical objects is modelled by the exchange of soft gravitons $h_{\mu\nu}$, defined as usual as a perturbation around the flat metric $\eta_{\mu\nu}$, i.e.\ $\kappa h_{\mu\nu}=g_{\mu\nu}-\eta_{\mu\nu}$, with $\kappa=\sqrt{32\pi G}$. 
Observables can be then derived, for example, from the eikonal phase $\chi$, which consists of resummed subsets of graviton exchanges. This is achieved very elegantly in the GWL approach \cite{Bonocore:2021qxh,Bonocore:2024uxk}, where it can be shown that subsets of graviton interactions relevant in the classical limit exponentiate, thus resumming the classical part of the scattering amplitude to all orders in the coupling $\kappa$. 

In particular, the higher-spin GWLs are defined through the path integral expression
\begin{equation}
    \widetilde{W}(p,\eta,b)=\int\mathcal{D}[x,\psi]\exp\left\{i\int_{-\infty}^{\infty}dt L[x,\psi]\right\},
\end{equation}
with the Lagrangian 
\begin{equation}\label{eq:lagrangianGWL}
    -\frac{\dot x^2}{2}+\frac{i}{2}\psi^a_i\frac{D}{dt}\psi_i^a-\frac{1}{2}h_{\mu\nu}(p+\dot x)^\mu(p+\dot x)^\nu+\frac{1}{2}(m^2-\mathfrak{M}^2).
\end{equation}
Here, $\mathfrak{M}$ is the dynamical mass in configuration space evaluated at position 
$b+pt+x$, where $b$ and $p$ refer to the impact parameter and momentum of the particle, respectively. The boundary conditions for the Grassmann fields $\psi_i^\mu(-\infty)+\psi^\mu_i(\infty)=2\eta_i^\mu$ generate the classical spin tensor $\cS^{\mu\nu}=i\eta^\mu_i\eta^\nu_i$. The Green functions for worldline fields and the procedure for solving the path integral have been discussed thoroughly in \cite{Bonocore:2021qxh,Bonocore:2024uxk}.

The eikonal phase $\chi$ can be computed from the vacuum expectation value of two GWLs w.r.t.\ the Einstein-Hilbert action
$S_{\rm{E.H.}}$, i.e.
\begin{equation}\label{eq:eikonaldef}
   e^{i\chi}=\int \mathcal{D}h_{\mu\nu}\widetilde{W}(p_1,\eta_1,b_1)\widetilde{W}(p_2,\eta_2,b_2)e^{iS_{\rm{E.H.}}},
\end{equation}
which results in a diagrammatic expansion in terms of the classical amplitudes that make up the GWL (i.e.\ three-point, Compton and higher-point amplitudes). 
The perturbative PM expansion of observables follows from the expansion of the Lagrangian in \cref{eq:lagrangianGWL} to the corresponding order. Using the Hamiltonians from the previous section we can compute two distinct GWLs: the first one in $d=4$ to all order in spin and limited to 1PM observables, and the second one to cubic order in spin, generic $d$ and suited for 2PM observables.  

\subsection{GWL at 1PM, all orders in spin and $d=4$}
Computing the 1PM Wilson line requires no propagators of worldline fields. Linearizing the all order in spin Hamiltonian in $\kappa$ and selecting the covariant SSC for the background parameters $p_\mu \eta^\mu_i=0$ yields 
\begin{align}\label{eq:GWLallspin}
	&\nonumber\widetilde{W}(p,\eta,b)=\exp\Bigg\{\nonumber\left(-\frac{i\kappa m}{2}\right)\int_k\tilde{h}_{\mu\nu}(k) e^{-ikb}\hat{\delta}(vk)\\&\Big(v^\mu v^\nu \cosh(\mathfrak{a}\cdot k)+i\frac{v^{(\mu}\mathcal{S}^{\nu)\sigma}k_\sigma}{m (\mathfrak{a}\cdot k)}  \sinh(\mathfrak{a}\cdot k)\Big)\Bigg\},
\end{align}
where $v^\mu=\frac{1}{m}p^\mu$ and $\mathfrak{a}_\mu=\frac{1}{2}v^\nu\epsilon_{\nu\mu \varrho\sigma}\cS^{\varrho\sigma}$. The exponent encapsulates the on-shell Kerr black hole three-point amplitude \cite{Guevara:2018wpp,Chung:2018kqs}, while off-shell terms were omitted. This amplitude can also be obtained by considering a generalization of minimal coupling for higher-spin massive particles \cite{Arkani-Hamed:2017jhn}, offering a new perspective on the notion of minimal coupling on the worldline. Specifically, the Kerr limit singles out a class of couplings on the worldline via \cref{eq:Kerrcondition}, thus suggesting an analogue of minimal coupling for amplitudes on the supersymmetric worldline. 

Observables are obtained by differentiating the eikonal phase of \cref{eq:eikonaldef} w.r.t.\ $b^\mu$ or $\mathfrak{a}^\mu$, and they agree with the results in the literature \cite{Brandhuber:2023hhl,Guevara:2019fsj}. The all order in spin GWL is a new result. It explicitly demonstrates for the first time the exponentiation of the Kerr three-point amplitude. The non-triviality of this result can be appreciated from the perspective of the amplitude approach, where already the exponentiation of the linear-in-spin three-point amplitude is quite involved \cite{Haddad:2021znf}.

\subsection{GWL at 2PM, cubic order in spin and generic $d$}
At 2PM and cubic order in spin the GWL for generic $d$ takes the form
\begin{align}\label{eq:GWLcubicinspin}
	&\widetilde{W}(p,\eta,b)=\exp\Bigg\{-\frac{i m}{2}\Big(\kappa\int_k \tilde h_{\mu\nu} e^{-ikb}\hat{\delta}(kv) V_3^{\mu\nu}\\&+\nonumber\kappa^2 \int_{k,l} \frac{\tilde h_{\mu\nu}(k)\tilde h_{\varrho\sigma}(l)}{2} e^{-ib(k+l)}\hat \delta(v(k+l)) V_4^{\mu\nu\varrho\sigma}\Big)\Bigg\}.
\end{align} 
Here, $V_3^{\mu\nu}$ is the three-point amplitude, which reads
\begin{align}\label{}
	&V_3^{\mu\nu}\!=v^\mu v^\nu\!+\frac{i}{m}v^{(\mu}(\cS\!\cdot\! k)^{\nu)}\!\left(1+\frac{ C_{BS^3}}{6m^2}(k \!\cdot\! \cS\!\cdot\! \cS\! \cdot\! k)\!\right)\\&+\frac{\tilde C_{ES^2}}{2m^2} v^\mu v^\nu (k \!\cdot\! \cS\!\cdot\! \cS\! \cdot\! k)-\frac{C_{ES^2}+\tilde C_{ES^2}}{2m^2} (\cS\!\cdot\! k)^\mu(\cS\!\cdot\! k)^\nu\nonumber,
\end{align}
where we have again omitted off-shell contributions.
The Compton amplitude $V_4^{\mu\nu\varrho\sigma}$ can be written compactly by including the three-graviton vertex contribution and contracting with the graviton polarization vectors. Introducing the gauge invariant combinations $F_i^{\mu\nu}=k_i^\mu \epsilon^\nu_i-k_i^\nu \epsilon^\mu_i$, we find 
\begin{align}\label{eq:vi}
    &V_4^{\mu\nu\varrho\sigma}\epsilon_{1,\mu}\epsilon_{1,\nu}\epsilon_{2,\varrho}\epsilon_{2,\sigma}\\=&\nonumber2\frac{V_{(0)}+V_{(1)}+V_{(2)}+V_{(3)}}{(k_1+k_2)^2}\text{Re}\left(\frac{1}{(v(k_1-k_2)+i\epsilon)^2}\right),
\end{align}
where the terms $V_{(i)}$ are listed in \cref{eq:Compton}.

The eikonal phase can be constructed from the following diagrams in terms of the GWL amplitudes (omitting mirrored expressions)  
\begin{align}\label{}
	&i\chi=\begin{tikzpicture}[baseline=(mid)]
		\tikzset{blob/.style={draw=black,circle,/tikz/inner sep=5pt ,pattern color=black,pattern=north west lines}}
		\begin{feynman}
			\vertex (mid);
			\vertex [blob,above=0.3cm of mid] (tm) {};
			\vertex [blob,below=0.3cm of mid] (bm) {};
			\vertex [left=0.8cm of bm] (bl) ;
			\vertex [right=1cm of bm] (br) {} ;
			\vertex [left=0.8cm of tm] (tl) ;
			\vertex [right=1cm of tm] (tr) {};
			\diagram*{(bl)--[line width=1pt,double distance=1pt] (bm)--[line width=1pt,double 
				distance=1pt](br),(tl)--[line width=1pt,double distance=1pt] (tm)--[line width=1pt,double 
				distance=1pt](tr), (bm)--[photon](tm)};
		\end{feynman}
	\end{tikzpicture}+
	\begin{tikzpicture}[baseline=(mid)]
	\tikzset{blob/.style={draw=black,circle,/tikz/inner sep=5pt ,pattern color=black,pattern=north west lines}}
	\begin{feynman}
		\vertex (mid);
		\vertex [above=0.58cm of mid] (tm);
		\vertex [blob,below=0.3cm of mid] (bm) {};
		\vertex [left=1cm of bm] (bl) ;
		\vertex [right=1.6cm of bm] (br) {$p_2,\cS_2$} ;
		\vertex [blob,left=0.2cm of tm] (tl) {};
		\vertex [blob,right=0.2cm of tm] (tr) {};
		\vertex [left=0.6cm of tl] (tll) ;
		\vertex [right=1.2cm of tr] (trr) {$p_1,\cS_1$};
		\diagram*{(bl)--[line width=1pt,double distance=1pt] (bm)--[line width=1pt,double 
			distance=1pt](br),(tll)--[line width=1pt,double distance=1pt](tl)--[line width=1pt,double distance=1pt] (tm)--[line width=1pt,double 
			distance=1pt](tr)--[line width=1pt,double distance=1pt] (trr), (bm)--[photon](tl),(bm)--[photon](tr)};
	\end{feynman}\end{tikzpicture}.
\end{align}
We have checked the eikonal and scattering angle against available results \cite{Jakobsen:2021zvh,Bohnenblust:2024hkw,Bern:2022kto}.

In the GWL approach the exponentiation of the amplitudes is manifest, demonstrating that also the tidal dependence in the form of the coefficients $C_{ES^2},\tilde C_{ES^2}$, and $C_{BS^3}$ exponentiates (though $\tilde C_{BS^3}$ does not appear at this order in spin and $G$).

\section{Conclusions}
In this letter we have proposed non-minimal coupling to gravity for supersymmetric worldline models
as a promising framework 
to access higher-spin effects in the binary dynamics of black holes and neutron stars. In search for a suitable higher-spin Hamiltonian, we have deformed the supersymmetry charge and derived the necessary equation (\cref{eq:constraintT}) that enforces supersymmetry. We then solved this equation by truncating the expansion either at a certain order in the Riemann tensor or a certain order in the spin tensor. Both of these paths have proven to be a constructive way to build higher-spin Hamiltonians that preserve the spin supplementary condition. 

At linear order in the curvature, we have shown that there exists a unique solution in $d=4$ to all orders in spin (\cref{eq:dynamicalmass_allorders}).
This new result captures all orders in the mass dipole, leaving supersymmetry manifest. At leading order in the mass dipole, it reduces to the well-known Hamiltonian of Kerr black holes expressed in terms of the spin vector \cite{Levi:2015msa,Scheopner:2023rzp}. 
Moreover, the way that the Hamiltonian emerges from the supersymmetry algebra provides a new perspective on the notion of minimal coupling at higher orders 
 in spin, in parallel to on-shell methods \cite{Arkani-Hamed:2017jhn,Guevara:2018wpp,Chung:2018kqs}.
 
At cubic order in spin, we have provided for the first time the Hamiltonian for generic $d$ with full tidal dependence (\cref{eq:Hamiltonian_d_wilsoncoefficients}). In particular, we find two new Wilson coefficients and tensor structures entering the Hamiltonian, while in the $d\rightarrow4$ limit we recover the known tidal operators with coupling $C_{ES^2}$ and $C_{BS^3}$.

For both Hamiltonians, \cref{eq:Kerrcondition} sets all free Wilson coefficients to unity in $d=4$, hence singling out the Kerr geometry from a generic compact object. In this way the supersymmetric model provides a mechanism to obtain the Kerr geometry without any prior knowledge about this specific solution to Einstein's equation. Equipped with the two Hamiltonians and using the GWL approach, we demonstrated for the first time the explicit exponentiation of the Kerr three-point amplitude and the three-point and Compton amplitudes for generic compact objects  (\cref{eq:GWLallspin,eq:GWLcubicinspin}).

We believe the synergy between supersymmetry and non-minimal coupling provides a promising framework for increasing the theoretical accuracy of higher-spin gravitational wave observables. 
For instance, it paves the way for obtaining the still unknown terms beyond quadratic order in spin at 3PM. 
We also expect the all order in spin solution to generalize to higher orders in the Riemann tensor. This would provide new insights into the dynamical multipole moments of Kerr black holes, which are under active investigation \cite{Bern:2022kto,Aoude:2022trd,Aoude:2023vdk,Cangemi:2022bew} and contribute to the higher-spin Compton amplitude.

\begin{acknowledgements}
This work was supported by the Deutsche Forschungsgemeinschaft (DFG) through the
Research Training Group ”GRK 2149: Strong and Weak Interactions – from Hadrons to
Dark Matter” and the Excellence Cluster ORIGINS under Grant
No.EXC-2094-390783311. JP's work is also supported by 
the "DAAD - Forschungsstipendium für Doktorand*innen". JP would like to thank Prof. Radu Roiban and the Penn State Center for Spacetime and Geometry for hospitality during final stages of this work. 
\end{acknowledgements}

\appendix 

\begin{widetext}
\section*{Supplementary material}\label{app:compton}

We list here the terms appearing in \cref{eq:vi}.
\begin{subequations}\label{eq:Compton}
\begin{align}
 &V_{(0)}=2 \left(v\cdot F_1\cdot F_2\cdot v\right)^2\\
&V_{(1)}=i \left(v\cdot F_1\cdot F_2\cdot v\right)\left(k_1v (F_1\cdot F_2)^{\mu\nu}+(k_2\cdot F_1\cdot v)F_2^{\mu\nu}+(1\leftrightarrow 2)\right)S_{\mu\nu}\\
    &V_{(2)}=\nonumber-\frac{1}{2}\Big((k_1\cdot F_2\cdot v)(k_2\cdot F_1\cdot v)F_1^{\mu\nu} \cS_{\mu\nu}F_2^{\varrho\sigma} \cS_{\varrho\sigma}+F_1^{\mu\nu} \cS_{\mu\nu}(v\cdot F_1\cdot F_2\cdot k_1)(v\cdot F_1 \cdot \cS\cdot k_1)\\&+F_2^{\mu\nu} \cS_{\mu\nu}(v\cdot F_2\cdot F_1\cdot k_2)(v\cdot F_2 \cdot \cS\cdot k_2)\Big)-2\tilde C_{ES^2} (v\cdot F_1\cdot F_2\cdot v)(k_1\cdot k_2)(v\cdot F_1\cdot \cS \cdot \cS \cdot F_2\cdot v)\nonumber\\
    &+\nonumber(C_{ES^2}+\tilde C_{ES^2})\left(-(k_2\cdot F_1\cdot v)(v\cdot F_1\cdot \cS \cdot F_2\cdot \cS\cdot F_2 \cdot k_1)-(k_1\cdot F_2\cdot v)(v\cdot F_2\cdot \cS \cdot F_1\cdot \cS\cdot F_1 \cdot k_2)\right)\\&+\nonumber\frac{C_{ES^2}+\tilde C_{ES^2}}{4}\Big(-(k_1\cdot F_2\cdot v)^2(F_1^{\mu\nu} \cS_{\mu\nu})^2-(k_2\cdot F_1\cdot v)^2(F_2^{\mu\nu} \cS_{\mu\nu})^2+4(v\cdot F_1\cdot F_2\cdot v)(k_1+k_2)\cdot \cS \cdot \cS \cdot (k_1+k_2)\\&-4 (k_1\cdot  \cS\cdot k_2)(F_1\cdot F_2)^{\mu\nu} \cS_{\nu\mu}\Big)
    +\frac{C_{ES^2}+\tilde C_{ES^2}}{2}\Big(-(k_1\cdot F_2 \cdot v)F_1^{\mu\nu} \cS_{\mu\nu}((v\cdot F_1\cdot F_2 \cdot \cS \cdot k_1)+(v\cdot F_2\cdot F_1\cdot \cS\cdot k_2)\nonumber\\&-(v\cdot F_1\cdot \cS \cdot F_2 \cdot k_1))+ (k_1\cdot F_2 \cdot v)(v \cdot F_1 \cdot \cS \cdot k_2)(F_1\cdot F_2)^{\mu\nu} \cS_{\nu\mu}-(v\cdot F_1 \cdot \cS \cdot F_2 \cdot k_1)^2\nonumber\\&+(v \cdot F_1 \cdot F_2 \cdot k_1)(v\cdot F_1 \cdot \cS \cdot F_2 \cdot \cS \cdot k_1)+F_1^{\mu\nu} \cS_{\nu\mu}(v\cdot F_2\cdot F_1 \cdot k_2)(v\cdot F_2 \cdot \cS \cdot k_2)+(1\leftrightarrow 2)\Big)\\
    &V_{(3)}=\nonumber-i\tilde C_{ES^2}\Big((k_2\cdot F_1\cdot v)(k_1\cdot \cS\cdot F_1\cdot k_2)(v\cdot F_2 \cdot \cS\cdot\cS\cdot F_2 \cdot v)\\&\nonumber
    +2(k_2\cdot F_1\cdot v)(v\cdot F_1 \cdot \cS\cdot k_1)(k_1\cdot F_2\cdot \cS\cdot \cS\cdot F_2 \cdot v)+2(v\cdot F_1\cdot F_2\cdot k_1)(v\cdot F_1 \cdot \cS\cdot k_1)(v\cdot F_2\cdot \cS\cdot \cS\cdot k_2)\Big)\\&\nonumber
    +i\tilde C_{ES^2}^2\Big((v\cdot F_1\cdot F_2 \cdot v)(k_1\cdot k_2)(v\cdot F_2\cdot \cS^3\cdot F_1\cdot v)+(k_1\cdot k_2)(v\cdot k_1)(v\cdot F_2 \cdot\cS^2\cdot F_1\cdot v)(v\cdot F_1\cdot\cS\cdot F_2\cdot v)\Big)\\&\nonumber+i2\tilde C_{ES^2}(C_{ES^2}+\tilde C_{ES^2})\Big(-(v\cdot F_1\cdot F_2\cdot k_1)(v\cdot F_1\cdot\cS\cdot k_1)(v\cdot F_2\cdot\cS^2\cdot k_1)\\&\nonumber+(k_2\cdot F_1\cdot v)(k_1\cdot F_2\cdot v)(k_1\cdot\cS\cdot F_1\cdot\cS^2\cdot F_2 \cdot v)+(k_2\cdot F_1\cdot v)(k_2\cdot \cS\cdot F_2 \cdot k_1)(v\cdot F_1\cdot\cS^2\cdot F_2 \cdot v)\Big)\\&\nonumber-
    \frac{i}{4}(C_{ES^2}+\tilde C_{ES^2})(k_2\cdot F_1 \cdot v)(k_2\cdot\cS\cdot F_2\cdot k_1)(F_1^{\mu\nu}\cS_{\mu\nu})(F_2^{\mu\nu}\cS_{\mu\nu})\\&\nonumber-
    i(C_{ES^2}+\tilde C_{ES^2}-1)^2(v\cdot F_2\cdot \cS\cdot k_2)(v\cdot F_1\cdot\cS^2\cdot F_2\cdot v)(k_1\cdot \cS\cdot F_1\cdot k_2)\\&\nonumber-
    \frac{i}{4}(C_{ES^2}+\tilde C_{ES^2})^2\Big(2(k_1\cdot F_2 \cdot v)(k_2\cdot \cS\cdot F_2 \cdot \cS\cdot F_1 \cdot k_2)(F_1^{\mu\nu}\cS_{\mu\nu})\\&\nonumber+2(k_1\cdot\cS\cdot F_1\cdot k_2)(v\cdot F_1\cdot \cS\cdot F_2 \cdot k_1)(F_2^{\mu\nu}\cS_{\mu\nu})+(k_1\cdot\cS\cdot k_2)(v\cdot F_1\cdot F_2\cdot k_1)(F_1^{\mu\nu}\cS_{\mu\nu})(F_2^{\mu\nu}\cS_{\mu\nu})\Big)\\&\nonumber-
    \frac{i}{3}C_{BS^3}(k_1\cdot F_2\cdot v)\Big((v\cdot F_2\cdot \cS\cdot q_2)(v\cdot F_1\cdot\cS^2\cdot F_1\cdot k_1)+(v\cdot F_1\cdot \cS\cdot k_1)(v\cdot F_1\cdot F_2\cdot\cS^2\cdot (k_1+k_2))\\&\nonumber+(v\cdot F_1\cdot\cS\cdot k_2)(v\cdot F_1\cdot F_2\cdot\cS^2 \cdot k_2)+(k_2\cdot\cS\cdot F_1\cdot k_2)(v\cdot F_2\cdot\cS^2\cdot F_1 \cdot v)\\&\nonumber+(k_1\cdot\cS\cdot k_2)(v\cdot F_1\cdot\cS^2\cdot F_2\cdot F_1 \cdot v)+(k_2\cdot F_1\cdot v)(v\cdot\cS\cdot (F_2\cdot \cS+\cS\cdot F_2)\cdot\cS\cdot (k_1+k_2))\\&\nonumber+(v\cdot F_1\cdot F_2 \cdot v)((k_1\cdot \cS\cdot F_1 \cdot\cS^2 \cdot (k_1+k_2))+(k_1\cdot\cS^2 \cdot F_1 \cdot \cS\cdot k_2))-2(v\cdot F_1 \cdot \cS\cdot k_2)(v\cdot F_1\cdot F_2 \cdot\cS^2\cdot k_1)\Big)\\&\nonumber
    +\frac{i}{3}C_{BS^3}\Big((k_1\cdot k_2)(v\cdot F_2\cdot \cS\cdot F_1 \cdot k_2)(v\cdot F_2 \cdot\cS^2\cdot F_1\cdot v)+(k_1\cdot\cS\cdot k_2)(v\cdot F_2\cdot F_1\cdot k_2)(v\cdot F_1\cdot\cS^2\cdot F_2\cdot v)\\&\nonumber-(k_1\cdot\cS^2\cdot k_2)(v\cdot F_2\cdot F_1\cdot k_2)(v\cdot F_1\cdot \cS\cdot F_2\cdot v)-(k_2\cdot\cS^2\cdot k_2)(v\cdot F_2\cdot F_1\cdot k_2)(v\cdot F_1\cdot\cS\cdot F_2\cdot v)\\&\nonumber+(k_1\cdot\cS\cdot k_2)(v\cdot F_1\cdot F_2\cdot v)((k_1+k_2)\cdot\cS^2\cdot F_2\cdot F_1\cdot v)\\&-2(v\cdot F_2\cdot F_1\cdot k_2)(v\cdot F_2 \cdot \cS\cdot k_2)(v\cdot F_1\cdot\cS^2\cdot(k_1+k_2))\Big)
    +(1\leftrightarrow 2)
\end{align}
\end{subequations}
We note that the way to write the Compton amplitude in terms of contractions of the $F_i$-tensors is not unique. The Compton amplitude is on-shell and therefore the number of spacetime dimensions $d$ does only enter implicitly through the various $d$-dimensional tensors.
\end{widetext}
\newpage

\bibliography{ref}

\end{document}